\documentclass[aps,prb,reprint,superscriptaddress,floatfix,bibnotes]{revtex4-1} 
\usepackage{epstopdf}
\usepackage{graphicx} 
\usepackage{amsmath,amssymb}

\begin{document}

\title{Electron Energy Loss Function of Silicene and Germanene Multilayers on Silver}

\author{L.\ Rast}
\email[Electronic mail: ]{annalauren.rast@nist.gov}
\affiliation{Applied Chemicals and Materials Division, National Institute of Standards and Technology, Boulder, CO 80305}

\author{V.\ K.\ Tewary}
\affiliation{Applied Chemicals and Materials Division, National Institute of Standards and Technology, Boulder, CO 80305}


\begin{abstract}
We calculate electron energy loss spectra (EELS) for composite plasmonic structures based on silicene and germanene. A continued-fraction expression for the effective dielectric function is used to perform multiscale calculations of EELS for both silicene and germanene-based structures on silver substrates. A distinctive change in plasmonic response occurs for structures with a germanene or silicene surface coating of more than three layers. These differences may be exploited using spectroscopic characterization in order to determine if a few-layer coating has been successfully fabricated.
\end{abstract}
\maketitle

\section{Introduction}
Silicene and germanene, two dimensional allotropes of silicon and germanium, have recently attracted attention as two-dimensional materials beyond graphene.~\citep{Bechstedt2012, Chinnathambi‎2012, C3CP51078F, C3CP51028J, Scalise2013,doi:10.1021/nl203065e} These materials possess predicted electron transport properties similar to graphene,~\citep{Tsai2013} as well as the advantage of compatibility with existing silicon-based technology. Additionally, the inversion symmetry breaking imparted by the buckled lattice structure of both materials may be taken advantage of through the application of an external electrical field perpendicular to the plane for highly controllable band gap tunability.~\citep{PhysRevLett.107.076802,PhysRevB.84.195430,Tsai2013,doi:10.1021/nl203065e}

The reactivity of silicene and germanene mean that they are more challenging to fabricate than graphene. Both materials bond easily with other materials and may oxidize rapidly in air, and are therefore fabricated using techniques such as epitaxial growth under ultra-high vacuum.~\cite{PhysRevLett.108.155501} Electron energy loss spectroscopy (EELS) and optical absorption spectroscopy are convenient and broadly used materials characterization techniques. We demonstrate that distinctive differences in plasmonic response open up the possibility for the use of EELS or absorption spectroscopy to distinguish few-layer coating silicene or germanene on silver from bare samples as well as bulk coatings. 

Determination of the collective opto-electronic properties of composite multilayer structures, particularly those based on two dimensional materials, necessitate realistic treatment of both the individual material properties and the interactions among the constituent materials.~\citep{PhysRevB.87.045428} We detail and employ such a method for the calculation of electron energy loss spectra (EELS) of multilayer structures consisting of silicene and germanene layers on silver and silver/silicon substrates.

This effective dielectric function is based on a specular reflection model, first derived by Lambin et al.~\cite{Lambin1985}, and takes into account the boundary conditions across each layer in the stratified structure. The use of this efficient continued-fraction expression along with pre-prepared libraries of dielectric functions for the individual materials allows for extremely efficient calculation of a wide variety of configurations for multilayer composites.

\section{EELS Calculation Details}
\subsection{General Procedure}
\begin{figure}
\includegraphics[scale=1.0]{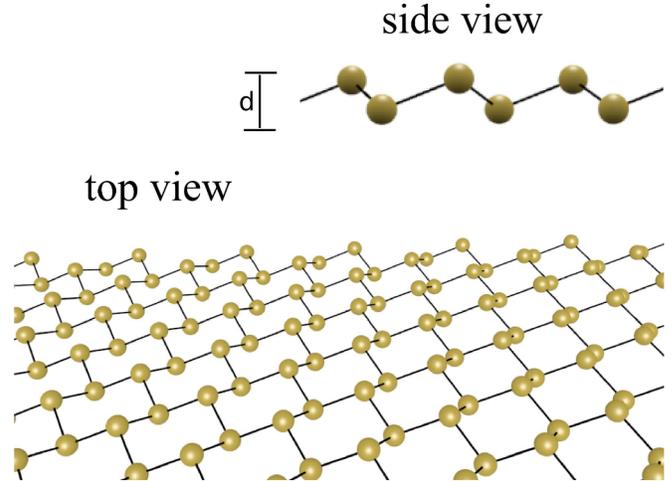}
\caption{\label{silgerstruct}Top and side views of monolayer crystal structures for (a) silicene and (b) germanene. Buckling amplitude d, obtained from the literature, is $.44\,$\AA for silicene and $.6\,$\AA for germanene.}
\end{figure}

\begin{figure}
\includegraphics[scale=1.0]{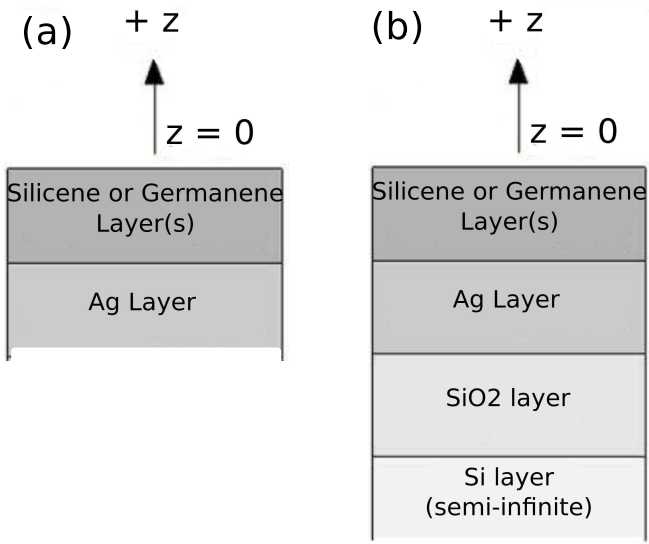}
\caption{\label{Multilayer}Multilayer structure: EELS are calculated for a variety of structures consisting of (a) silicene and germanene top layer(s) and semi-infinite Ag substrate and (b) silicene and germanene top layer(s), Ag middle layer, and SiO$_2$/Si substrate.}
\end{figure}

Individual complex dielectric functions are first obtained for each layer.  The crystal structures used for the individual layer dielectric functions are depicted in ~\ref{silgerstruct}. Then, EELS are calculated for a variety of multilayer sandwich structures as depicted in Fig.~\ref{Multilayer}. We calculate the values of the silicene and germanene dielectric functions through ab-initio density functional theory (DFT) methods including excitonic effects. The buckling amplitude d, obtained from DFT structural calculations in~\citep{Scalise2013}, is $.44\,$\AA for silicene and $.6\,$\AA for germanene. These values are in good agreement with the values obtained by others.~\citep{C3CP51078F} Empirical values from the literature are used for the silver and silicon substrate layers. These values are then stored for use as input to a continued-fraction algorithm, which yields the effective dielectric function. This algorithm is outlined in Section~\ref{composite}.

\subsection{\label{composite}The Effective Dielectric Function}
As discussed in our previous work on graphene,~\citep{PhysRevB.87.045428} the effective dielectric function, $\xi(\omega,k,z)$, of the stratified structure in Fig.~\ref{Multilayer} is that of Lambin et al.~\cite{Lambin1985} The expression for $\xi$ was derived from EELS theory in a reflection geometry. This expression has been shown to be applicable to both phonons~\cite{Lambin1985} and polaritons~\cite{PhysRevB.38.5438} in stratified structures with histogram-like dielectric functions (continuous within each layer) and interacting interfaces. The expression and the formalism from which it is derived were first appplied to describing composite dielectric functions semiconducting materials, but are also applicable to surface plasmon resonance and phonon behavior of alternating of metal-insulator layers.~\cite{PhysRevB.38.5438,PhysRev.182.539} This model has been shown to be in good agreement with the well-known Bloch hydrodynamic model in the small wave vector regime considered in this work.~\cite{Ritchie1966234}

The $z$ coordinate is in the direction perpendicular to the free surface of the sample, extending from the $z = 0$ surface to $-\infty$. $\mathbf{k}$ denotes the surface excitation (plasmon or phonon) wave vector and $\omega$ is the frequency of excitation.
 
\begin{equation}
\xi (\mathbf{k},\omega,z) = \frac{i\mathbf{D}(\mathbf{k},\omega,z)\cdot \mathbf{n}}{\mathbf{E}(\mathbf{k},\omega,z)\cdot\mathbf{k}/k},
\end{equation}
where $\mathbf{D}(\mathbf{k},\omega,z) = \epsilon(\omega,z)\mathbf{E}(\mathbf{k},\omega,z)$, and $\epsilon(\omega,z$) is the long wavelength dielectric function (tensor) of the material at $z$. $\xi$ remains continuous even in the case of sharp interfaces parallel to the $x$-$y$ directions below the surface (as is the case in our multilayer system). This is due to the interface boundary conditions: continuity of $D_\perp$ and $E_\parallel$. 

The effective dielectric function $\xi_0(k,\omega)$ (Eq.~\ref{continuedfrac}) is a solution to the Riccati equation (Eq.~\ref{ricatti}), in the long-wavelength approximation $k\approx0$, at the $z=0$ surface.~\cite{Lambin1985} We fix $k$ as $k = 0.005\,$\AA$^{-1}$ for both the ab-initio calculations and the composite calculation. 
Eq.~\ref{ricatti} was derived for heterogeneous materials made of a succession of layers (with homogeneous dielectric functions within each layer), the layers having parallel interfaces. $\epsilon(z)$ are complex functions, with positive imaginary parts at $z=0$.~\cite{Lambin1985} 

\begin{eqnarray}\label{ricatti}
\frac{1}{k}\frac{\mathrm{d}\xi(z)}{\mathrm{d}z} + \frac{\xi^2(z)}{\epsilon(z)} = \epsilon(z)
\end{eqnarray}
\begin{equation}\label{continuedfrac}
\xi_0 = a_1 - \frac{b_1^2}{a_1+a_2-\frac{b_2^2}{a_2+a_3-\frac{b_3^2}{a_3+a_4-\cdots }}}
\end{equation}
where 
\begin{eqnarray}
a_i=\epsilon_i\coth(kd_i)
\end{eqnarray}
and 
\begin{eqnarray}
b_i=\epsilon_i / \sinh(kd_i).
\end{eqnarray}

Once individual dielectric functions are obtained, this procedure allows for the performance of mesoscale EELS calculations of a wide variety of layered structures. Layer thickness and material are easily substituted, with each EELS calculation running in a less than a second on a single processor (nearly independent of the spectral range). EELS are calculated directly from the effective dielectric function as
\begin{equation}\label{EELS}
\mathrm{EELS} = \mathrm{Im}\left[\frac{-1}{\xi(\omega,k) + 1}\right].
\end{equation}

Inspection of Eq.~\ref{continuedfrac} reveals that for $\mathrm{Im}[\epsilon_i] > 0$, $\mathrm{Im}[\xi_0]>0$. EELS spectra given by Eq.~\ref{EELS} are then generally positive-valued.

\subsection{Silver Dielectric Function}
The silver dielectric functions are empirical values by Johnson and Christy~\cite{Johnson1972} obtained by reflection and transmission spectroscopy on vacuum-evaporated films at room temperature. Film-thickness in the Johnson and Christy study ranged from $185\,$\AA -- $500\,$\AA. It was found that optical constants in the film-thickness range $250\,$\AA\ -- $500\,$\AA\ did not vary appreciably. As in our previous work,~\citep{PhysRevB.87.045428} $340\,$\AA\ film thickness is representative of bulk mode dominant (yet still nanoscale) metallic thin films. 
\subsection{SiO$_2$ and Si Dielectric Constants}
Relative static permittivities of 3.9 and 11.68 were chosen for the SiO$_2$ and Si dielectric constants, respectively. These are reasonable and widely-used values obtained from the literature.~\cite{muraka2003, yi2012}
\subsection{Silicene and Germanene Individual Layer Dielectric Functions}
\begin{figure}
\includegraphics[scale=1.0]{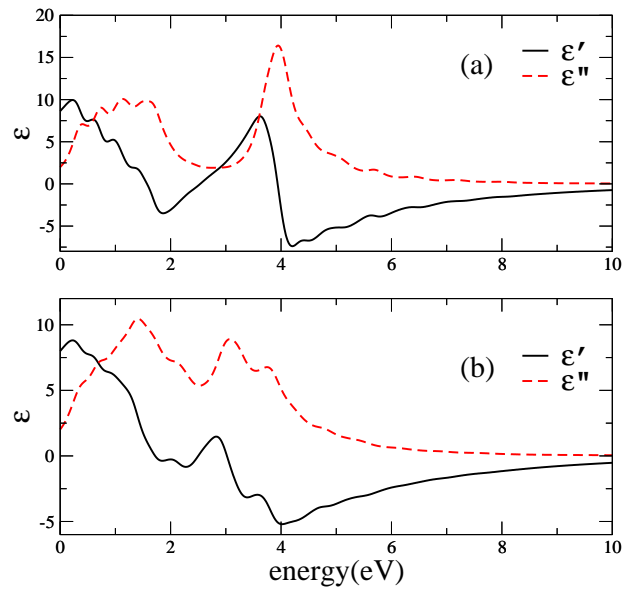}
\caption{\label{silger} Complex relative dielectric function $\epsilon(\omega)$ for silicene (a) and germanene (b). Real and imaginary parts ($\epsilon'(\omega)$ and $\epsilon''(\omega)$) are represented by solid and dotted lines, respectively.}
\end{figure}

Complex dielectric functions for silcene and germanene are displayed in Fig.~\ref{silger} (a) and Fig.~\ref{silger} (b), respectively. These ab-initio calculations use the time-dependent DFT with a GLLBSC exchange correlation functional,~\citep{PhysRevB.82.115106} and are implemented in the Python code GPAW, a real-space electronic structure code using the projector augmented wave method.~\footnote{Certain commercial equipment, instruments, or materials are identified in this paper in order to specify the experimental procedure adequately. Such identification is not intended to imply recommendation or endorsement by the National Institute of Standards and Technology, nor is it intended to imply that the materials or equipment identified are necessarily the best available for the purpose.}$^{,}$~\cite{gpaw1, gpaw2, gpaw3, gpaw4} Both silicene and germanene dielectric functions are calculated in the optical limit with a momentum transfer value of $0.005\,$\AA$^{-1}$, along the $\bar{\Gamma}$-$\bar{M}$ direction of the surface Brillouin zone.  The $k$-point sampling with $20 \times 20 \times 1$ Monkhorst--Pack grid was chosen for the band-structure and EELS calculations for both silicene and germanene. We have chosen to employ both the GLLBSC functional and the Bethe Salpeter Equation (BSE) in order to calculate the individual layer dielectric functions due to the extreme accuracy of this method in predicting experimental values of dielectric functions and bandgaps for similar materials, such as a variety bulk semiconductors including silicon as well two dimensional materials graphene and hexagonal boron nitride.~\cite{PhysRevB.86.045208}The GLLBSC potential explicitly
includes the derivative discontinuity of the xc-potential at integer particle numbers, critical for obtaining physically meaningful band structure via a DFT calculation. This functional has also been shown to have computational cost similar to the Local Density Approximation (LDA) with accuracy similar to methods such as the LDA-GW method.~\cite{PhysRevB.86.045208,PhysRevB.84.235430} The use of the BSE is important due to the inclusion of excitonic effects, an prominent spectral feature for both materials.~\citep{C3CP51078F} A two dimensional Coulomb cutoff ~\citep{PhysRevB.73.205119} is employed in order to calculate the diectric function of the silicene and germanene monolayers.

Our model utilizes dielectric functions due to surface parallel excitations only, as the effective dielectric function is derived in a specular reflection geometry. The dielectric functions we have obtained for silicene and germanene (see Fig.~\ref{silger} (a)) agree well with previous calculations in the literature ~\cite{Chinnathambi‎2012,Bechstedt2012}, and has particularly good agreement with the spectral profiles and peak positions in ~\cite{C3CP51078F}, where the authors used the BSE to include excitonic effects. 

\section{RESULTS}
\subsection{Silicene and Germanene on Silver/SiO2/Si Substrates: Varying the Noble Metal Layer Thickness }
\begin{figure}
\includegraphics[scale=.9]{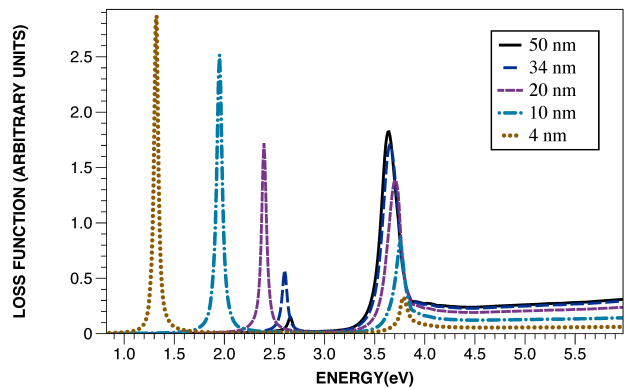}
\caption{\label{silicenesilverthickness}EELS for single-layer silicene on Ag/SiO$_2$/Si substrate : The effect of differing thickness for the Ag layer is demonstrated. The Ag layer thicknesses are $500\,$\AA\ (solid line), $340\,$\AA\ (long dashes), $200\,$\AA\ (short dashes), $100\,$\AA\ (dash-dot), and $40\,$\AA\ (dotted line).}
\end{figure}

\begin{figure}
\includegraphics[scale=.9]{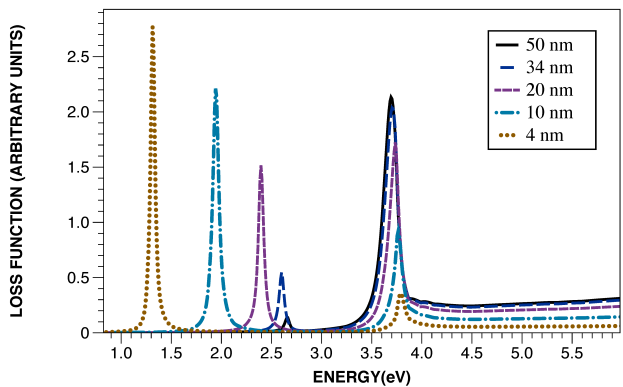}
\caption{\label{germanenesilverthickness}EELS for single-layer germanene on Ag/SiO$_2$/Si substrate : The effect of differing thickness for the Ag layer is demonstrated. The Ag layer thicknesses are $500\,$\AA\ (solid line), $340\,$\AA\ (long dashes), $200\,$\AA\ (short dashes), $100\,$\AA\ (dash-dot), and $40\,$\AA\ (dotted line).}
\end{figure}

Figures ~\ref{silicenesilverthickness} and ~\ref{germanenesilverthickness} demonstrate the effect of decreasing silver metallic layer thickness. As the Ag thickness is reduced, the so called \emph{begrenzung effect} is apparent. Enhanced surface-to-volume ratio in the metal causes stronger coupling to the surface resonance and decreased coupling to the bulk modes.~\cite{Ritchie1957, Osma1997} 

In the case of a thin metallic slab, empirical models have been thoroughly explored. Upon the introduction of a boundary to an infinite metallic slab, a negative (begrenzung) peak is introduced at the same energy as the bulk peak, and a trailing surface peak appears.~\cite{Ritchie1957} The surface peak becomes more intense with decreasing thickness, as does the negative begrenzung peak, decreasing the net bulk-plasmon amplitude. Surface modes become dominant for silver thickness between $20\,$ and $10\,$nm. This is consistent with observations in the well-validated and widely-used empirical data by Johnson and Christy~\cite{Johnson1972} as well as observations in our previous work on graphene/noble metal multilayer systems.~\cite{PhysRevB.87.045428} Comparison with experimental results for very thin silver layers provides further verification of the model for a wide variety of silver metal layer thicknesses. Both bulk and surface peak locations and relative intensities for $4\,$nm are in excellent agreement with experimental results for EELS of $3.4\,$nm silver layers.~\citep{Nagao2007}

\subsection{Silicene and Germanene Multilayers on Silver}
\begin{figure}
\includegraphics[scale=1.0]{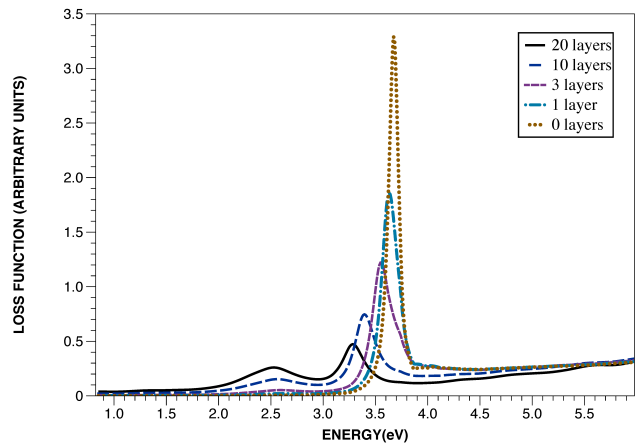}
\caption{\label{silicenelayers}EELS for multilayers of silicene on a semi-infinite Ag substrate : The effect of differing numbers of silicene layers is demonstrated. The silicene layer numbers are 20 (solid line), 10 (long dashes), 3 (short dashes), 1 (dash-dot), and 0 (dotted line).}
\end{figure}

\begin{figure}
\includegraphics[scale=.9]{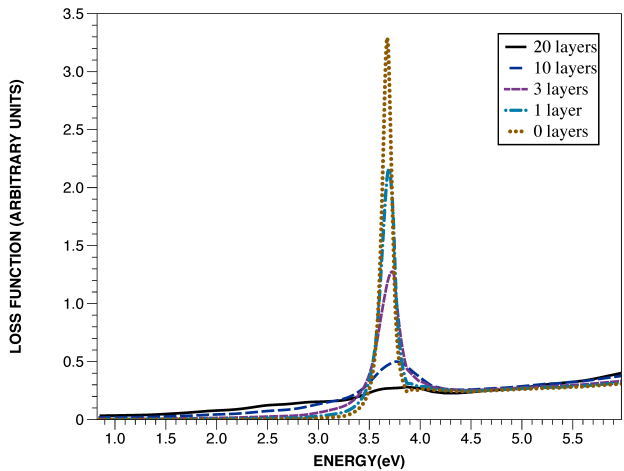}
\caption{\label{germanenelayers}EELS for multilayers of silicene on a semi-infinite Ag substrate : The effect of differing numbers of silicene layers is demonstrated. The silicene layer numbers are 20 (solid line), 10 (long dashes), 3 (short dashes), 1 (dash-dot), and 0 (dotted line).}
\end{figure}

Figures \ref{silicenelayers} and \ref{germanenelayers}demonstrate the effect of varying numbers of silicene and germanene layers on a silver substrate, respectively. For up to three layers of silicene on silver, the bulk plasmon peak is diminished without significant broadening. This indicates an overall reduction in bulk losses.  The effect is most notable when the silver slab is coated with a single layer of silicene, an effect which would be useful for determining successful fabrication of monolayer silicene on silver through spectroscopic characterization. At 10 layers and above, the system approaches the expected behavior for a bulk Si/Ag system, with a broad interfacial peak appearing at about 2.5 eV.~\citep{PhysRevB.87.045428} This peak broadens further and is enhanced in intensity with increasing number of silicene layers. Referring to Figure ~\ref{silger} (a), it is also apparent that at roughly 2.5 eV, the silicene dielectric function real part changes sign, and becomes increasingly positive up to nearly 4 eV.  The silver diectric function real part is very negative in this regime, so the interfacial plasmon is expected at this energy. This is in contrast to the germanene dielectric function, which is only momentarily slightly positive (Figure ~\ref{silger} (b)) in this regime. As a result, figure \ref{germanenelayers} demonstrates that there is no well-defined interfacial plasmon for the germanene/silver system. Damping of the silver bulk plasmon for a few layers of two dimensional material, however, occurs in a very similar manner for the germanene/silver and silicene/silver systems. As in the case of the silicene/silver system, for 1-3 layers of germanene on silver, the bulk plasmon is diminished to a great extent without significant broadening of the peak.  

\section{DISCUSSION}
In this study we investigated the effect of varying numbers of silicene and germanene layers on an Ag substrate. For mono-, bi-, and tri-layer coatings of both silicene and germanene, bulk plasmon modes are significantly diminished without significant broadening, which would correspond to increased plasmonic losses. The significance of this is two-fold: (1) This marked reduction in bulk peak intensity should be of use for characterization of few-layer silicene and germanene systems on silver, as a few layers of either material leads to a diminishing of the bulk plasmon peak without a significant broadening or shift in peak position. In the case of silicene, an additional interfacial peak occurs and is enhanced for more than 10 layers, an indication that the silicene coating is approaching bulk thickness. In the case of germanene, the silver bulk plasmon is quenched at 20 layers.  The obvious differences in behavior for uncoated, few-layer-coated systems, and and bulk coatings are useful as simple guidelines in the fabrication of these new materials.(2) The boundary physics for silicene and germanene, within the context of our mesoscopic model, is similar to our findings for graphene~\citep{PhysRevB.87.045428} --- The addition of a graphene boundary layer on the metallic surface reduces coupling of excitations to bulk plasmons through the begrenzung effect. The origin of the begrenzung effect is a reduction of the degrees of freedom for excitations, and thus further surface confinement comes at the expense of bulk oscillations, leading to reduced losses. 

The mesoscopic model used in these calculations has some limitations that merit discussion. Results of this study are valid in the long-wavelength limit for which the continued fraction expression by Lambin et al. was derived. Additionally, coupling between layers is classical (via boundary conditions), and as a result inter-layer hopping is neglected. However, at least in the case of bilayer silicene, it has been shown that inter-layer hopping can be neglected.~\citep{Rui2013}  It has been argued that the buckled silicene geometry, arising from mixing sp2 and sp3 hybridization, blocks interlayer hopping in bilayer silicene, thus preserving Dirac-type dispersion.~\citep{Rui2013} If this explanation is correct, the same argument may also apply to germanene bi-layers.

In future work we plan to incorporate the effect of lattice strain on the optical properties of the composite for two reasons: (1) strain engineering is expected to provide a further means of plasmon tuning,~\cite{ciammarella2010} and (2) due to inherent lattice mismatch even in systems with epitaxial growth, strain effects are generally of interest for accurate prediction of plasmonic features in two dimensional and quasi-two dimensional material-based heterostructures.

\begin{acknowledgments}
The authors would like to thank Katie Rice, Ann Chiaramonti Debay, and Alex Smolyanitsky for helpful discussions. This research was performed while the first author held a National Research Council Research Associateship Award at the National Institute of Standards and Technology. This work represents an official contribution of the National Institute of Standards and Technology and is not subject to copyright in the USA. 
\end{acknowledgments}

\providecommand{\noopsort}[1]{}\providecommand{\singleletter}[1]{#1}%

\bibliographystyle{aipauth4-1}
\bibliography{apspaper}

\end{document}